\documentclass[preprint,showpacs,preprintnumbers,amsmath,amssymb]{revtex4}

\usepackage{graphicx}
\usepackage{dcolumn}
\usepackage{bm}

\begin{document}

\preprint{APS/123-QED}

\title{Collaborative tagging as a tripartite network}

\author{R. Lambiotte}
\email{Renaud.Lambiotte@ulg.ac.be}

\author{M. Ausloos}
\email{Marcel.Ausloos@ulg.ac.be}

\affiliation{%
SUPRATECS, Universit\'e de Li\`ege, B5 Sart-Tilman, B-4000 Li\`ege, Belgium
}%

\date{09/07/2005}

\begin{abstract}
We describe online collaborative communities by tripartite networks, the nodes being
persons, items and tags. We introduce projection methods in order to uncover the structures of the networks, i.e. communities of users, genre families... 
 To do so, we focus on the correlations between the nodes, depending on their profiles, and 
 use percolation techniques that consist in removing less correlated links and observing the shaping of disconnected islands. The structuring of the network is visualised by using a tree representation. The notion of diversity in the system is also discussed.
\end{abstract}


\maketitle

\section{Introduction}

Recently, new kinds of websites have been dedicated to the sharing of people's habits and tastes, examples including their preferences in music, scientific articles, movies, websites... 
These sites allow members to upload from their own computer a library that characterises their habits in the corresponding topic (an iTunes music library for instance), and next to create a web page containing this list of items. Additionally, the website proposes the users to discover new content by comparing their taste with that of other users, thereby helping them discover new musics/books/websites... that should 
(statistically) fit their profile. 

This method rests on a feedback between the users and a central server, and is usually called collaborative filtering.
The emergence of these collaborative websites answers the needs of Internet users to retrieve useful and coherent informations from the millions of pages and data that form the Web. 
 Let us stress that the use of statistical methods in order to make coherent suggestions from a user profile is common in commercial websites, i.e. Amazon. The main particularities of collaborative systems are: (i)  their non-commercial purpose, even though the frontier with commercial companies is more and more vague (see for instance the acquisition of {\em del.icio.us} by {\em Yahoo} in November 2005); (ii)  their transparency, namely these sites are relatively open and do not hide the profiles of each user, contrary to Amazon for instance. 
 From a scientific point of view, this transparency opens perspectives  in order to perform large scale experiences (including thousands of people) on taste formation,  quantitative sociology, musicology... 
 The available data also suggest alternative methods in order to perform large scale classifications of music/science/internet. Those sub-divisions should be based on the intrinsic structure of the audience of the items.
 
In parallel with this sharing and statistical comparing of content, collaborative websites usually propose tagging possibilities. This process, called  "folksonomy" (short for "folk taxonomy") means that the websites allow users to publicly tag their shared content, the key point being that their tag is not only accessible to themselves, but also to the whole ensemble of users. For instance, in the case of music sharing habits, a group like {\em The Beatles} is described in different ways, i.e. {\em pop, 60s, britpop...}, that depend on the different backgrounds, tastes, music knowledge 
or {\em network of acquaintances}... of the users.

Both methods, i.e. collaborative filtering (CF) and collaborative tagging (CT) lead to complex networks  from which structures have to be extracted in order to deliver useful informations to users.  In this work, we  discuss methods  that lead to  the identification of \textit{a priori} unknown collective behaviours, and to a hierarchical representation of the network structuring.
To do so, we 
focus on empirical data extracted from websites specialised in music, e.g. {\em audioscrobbler.com} and {\em musicmobs.com}, and in scientific articles, i.e. {\em citeulike.com}.  
We  show that the tagging collaborative process leads to a tripartite network, i.e. a network with three different kinds of nodes (the users, the items and the tags) and where the links relate three nodes of different kinds. 
The next step of the analysis consists in projecting the tripartite network on lower order networks.
To do so, we evaluate the correlations between the items/tags, depending on their use. Filtering methods \cite{lambi}, i.e. percolation idea-based (PIB) methods, allow to uncover the collective behaviours.  The resulting hierarchical structure of the network leads to a statistical definition of the notion of {\em genre}, and draws a direct link between collaborative filtering and taxonomy. Finally, we discuss methods for measuring the diversity  of people \cite{lambi2}.

\section{Classification methods}

In this section, we give a short review  of the usual strategies that can be used in order to classify and organise content \cite{gold}, as well as their main differences. 

\subsubsection{Taxonomies}
Taxonomies include the Dewey Decimal classification for libraries, computer directory systems, the Linnean system of classifying living things... By construction, a taxonomy is hierarchical and exclusive. In these systems, each item is associated to one category, which belongs to a more general category, each category belonging to a more general one until the root of the tree is attained. For instance, the music artists  {\em Charlie Parker} and {\em Charles Mingus} can reasonably be classified in the categories {\em Bebop} and {\em Free Jazz}, both of them belonging to the category {\em Jazz}. By construction, taxonomies lead to an automatic structuring of content into hierarchical structures, that allow users to search with different levels of specificity. 

\subsubsection{Tagging systems}
Tagging systems are non-hierarchical and non-exclusive. They consist in associating to each item a list of keywords, all the keywords being considered at the same level. Tagging systems are especially adapted for  content that is not easily categorisable into exclusive categories, and for situations when no hierarchical difference exists between categories. Let us take the example of music. In addition to the usual genre classification of a music group, a listener may consider additional terms describing its mood, i.e. {\em Sad, Nervous, Happy}... A taxonomical description requires a hierarchical  organisation, i.e. a music group is placed in a directory $Jazz/Sad$ or in a directory $Sad/Jazz$. In a case when the importance of each characteristic is not clear, such a hierarchy is obviously not adequate, and may lead to problems in order to retrieve all relevant items. For instance, a music group placed in $Sad/Jazz$ is not found is the hieriarchically higher category $Jazz$.

\subsubsection{Collective description}
Usually, the choice of the set of tags available is done by an authority, such as a librarian or an editor, while the attribution of these tags is performed by the same authority or by
 the creators of the item, i.e. the authors of a scientific paper. It is only recently that websites have led to the emergence of collaborative tagging, also called folksonomy. 
Contrary to the usual tagging
classification systems, folksonomy is:

 (i) {\em anarchic}: the choice for the keywords is not restrained by any carcan (contrary to
PACS classifications in physics literature for instance), but may include any word composed of letters.

(ii) {\em democratic}: the tagging is equivalently performed by a large ensemble of persons, and not by a central one.

In itself, folksonomy is especially suitable for systems where no authority is present in order to organise the classifications. That is one of the reasons why it is gaining popularity on the web. The {\em democratic} aspect of the method also leads to a very rich description for each item. Namely items that are tagged by many persons 
are usually characterised by a spectrum of tags, revealing the diverse levels of descriptions associated to them. 
Nonetheless, the richness of the 
methods may also be a weakness in practice,  in order to retrieve useful information from a database for instance. This is due to the very large number of tags associated to each item, as well as to the use of terms that have not been {\em optimised} by an authority, such as synonyms or words written with several orthographs.

\subsubsection{Beyong the words}
Finally,  collaborative filtering (CF) is also a democratic method of classification, but, contrary to the above classifying methods, it does not require the use of words in order to attribute a category to items. Usually, CF uses statistical methods in order to link items depending on the people who use it. In the case of music, for instance, music groups are related if they have common audiences.  
Consequently, the observed categories rest on a more subtle description of items than the limited use of tagging words.
This is especially true in music where more and more groups prefer to be associated to {\em influential groups}  rather than to being categorised in usual subdivisions \cite{deus}. In a website like {\em www.myspace.com}, for instance, the attributes of an artist encompass such a list of influences.

\section{Methodology}

\subsection{Tripartite structure}

The structure of collaborative websites can be viewed as a tripartite network. Namely, it is  a network composed of three kinds of nodes: i) the persons or users $\mu$; ii)   the items  $i$ that can be music groups or scientific articles; iii) the tags $I$ that are used by the person $\mu$ to describe the item $i$. Depending on the systems under consideration, a person can use one or several tags on each item. 
The resulting network can be represented by a graph where edges run between the item $i$ and the user $\mu$, passing through the tag $I$. Moreover,  a weight is attributed to each link depending on the number of tags given by $\mu$ to $i$. For instance, if $\mu$ uses two tags for $i$, the weight of the links is $\frac{1}{2}$.

Let us note $n_U$ the number of users, $n_{It}$ the number of items, and $n_T$ the number of tags in the considered sample.
Consequently, each listener $\mu$ can be characterised by the $n_{It} \times n_T$ matrix $\overline{\overline{\sigma}}^\mu$:
\begin{equation}
\label{vector}
\overline{\overline{\sigma}}^\mu =  \left( \begin{array}{ccccccc}
0 & ...  &  1/2 & ... & 1/2 & ... & 0
\\
...  &  ... & ... & ... & ... & ... & ...
\\
...  &  1/3 & ... & 1/3 & ... & ... & 1/3
\\
...  &  ... & ... & ... & ... & ... & ...
\end{array} \right) ~~ 
\end{equation}
where $\overline{\overline{\sigma}}^\mu_{i I}$ denotes the weight of tag $I$ in its description of $i$, so that $\sum_I \overline{\overline{\sigma}}^\mu_{i I} = 1$ if $\mu$ owns $i$ and zero otherwise.
Each item and each tag is also characterised by similar matrices that we note $\overline{\overline{\gamma}}^i$ and $\overline{\overline{\alpha}}^I$ respectively.

\begin{figure}
\hspace{1.8cm}
\includegraphics[width=4.1in]{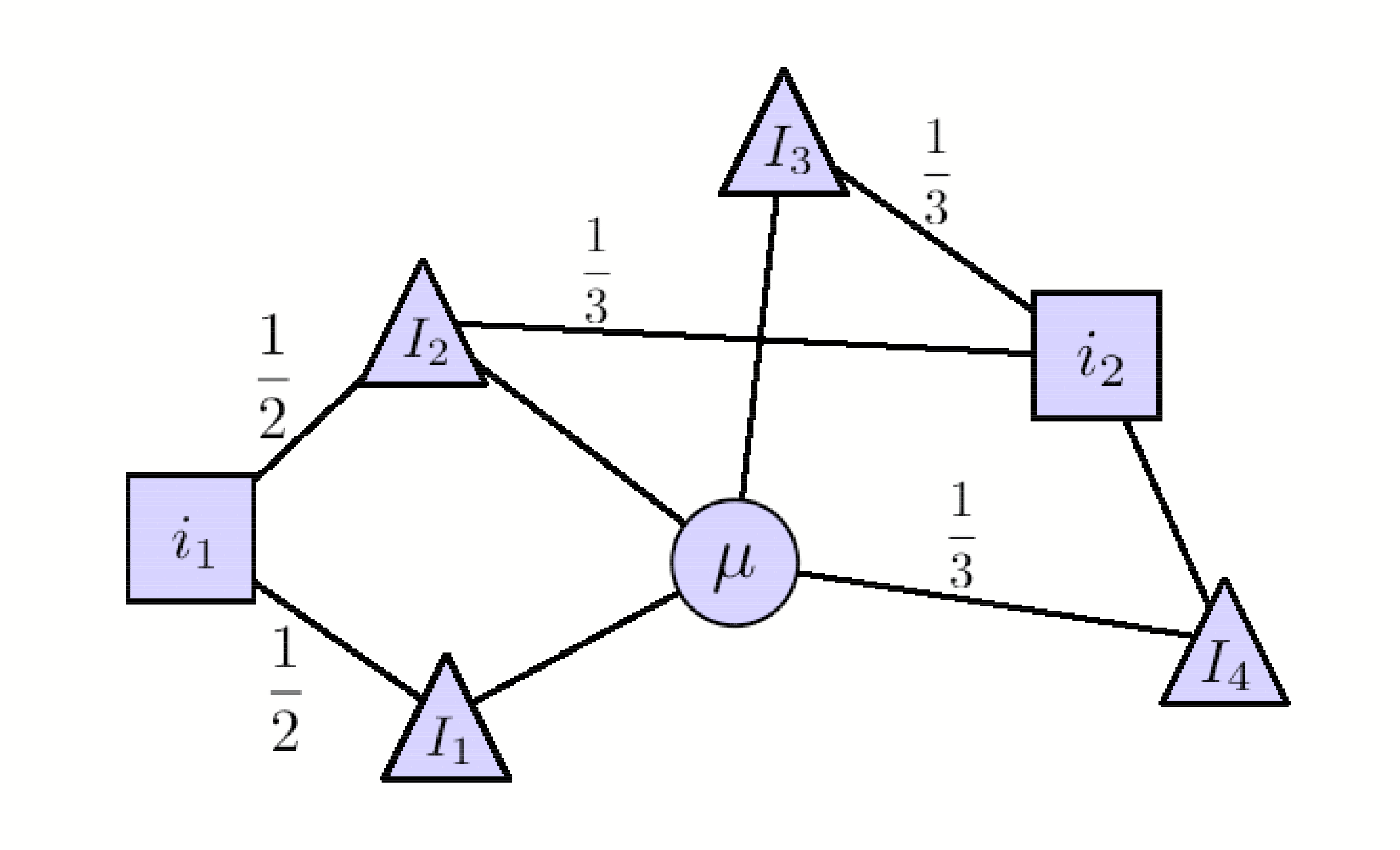}
\caption{\label{explication}  Tripartite structure of the tagging system. In this example, user $\mu$ owns two items $i_1$ and $i_2$ that are respectively tagged by two keywords, $(I_1,I_2)$, and three keywords $(I_2,I_3,I_4)$.}
\end{figure}

\subsection{Projecting method}

A common way to simplify the analysis of multi-partite networks consists in projecting them on lower order networks, i.e. unipartite or bipartite networks.

\subsubsection{Bipartite networks}
In the following, we  only focus on the correlations between two kinds of nodes, for instance between the users and the items. To do so, we first reduce the tripartite network to a bipartite one by summing over all nodes of one kind, thereby neglecting possible correlations between the three kinds of nodes. Such neglected correlations may include  the role of the specialisation of a user on the way he tags  items. For instance, a $Jazz$ lover will have a tendency to use more specific tags for $Jazz$ bands, because of i) his knowledge of more specific tags, and ii) his need of specialised tags  in order to retrieve informations from the many $Jazz$ songs composing his library. 

By using the above reduction method, the bipartite network users-item is obtained by summing over all tags, 
so that each listener $\mu$ is now described by the the $n_{It}$-vector $\overline{\sigma}^\mu_{|_I}$:
\begin{equation}
\label{vector}
\overline{\sigma}^\mu_{|_I} = (..., 1, ... , 0, ... ,1, ...),
\end{equation}
the index running over all items, and where $\overline{\sigma}^\mu_{|_I} = \sum_I \overline{\overline{\sigma}}^\mu_{i I}$.
The items are characterised by the $n_{U}$-vector $\overline{\gamma}^i_{|_I} = (..., 1, ... , 0, ... ,1, ...)$.
These vectors are signatures of the users/items, that account for their interests/audience. In the case of music, we call these vectors the {\em music signatures} of people and groups.
In the following, we also focus on the bipartite network item-tag, where the information about users has  been eliminated. It is accordingly defined by summing over all users, and leads to the vectors $\overline{\gamma}^i_{|_{\mu}}$ and $\overline{\alpha}^I_{|_{\mu}}$.

\subsubsection{Unipartite networks}
In order to project the bipartite network on a unipartite one, we look at the correlations between two nodes of the same kind, relatively to his behaviour with another kind. For instance, one may look how persons $\mu$ and $\lambda$ are correlated by using common items. To do so, we introduce the symmetric correlation measure:
\begin{equation}
\label{cosine}
C_{CF}^{\mu \lambda} = \frac{\overline{\sigma}^\mu_{|_I} . \overline{\sigma}^\lambda_{|_I}}{|\overline{\sigma}^\mu_{|_I}| |\overline{\sigma}^\lambda_{|_I}|} \equiv \cos \theta_{\mu \lambda} 
\end{equation}
where $\overline{\sigma}^\mu_{|_I} . \overline{\sigma}^\lambda_{|_I}$ denotes the scalar product between the two $n_{It}$-vector, and $||$ its  associated norm. This correlation measure, that corresponds to the cosine of the two vectors in the  $n_{It}$-dimensional space, vanishes when the persons have no common item, and is equal to $1$ when their item libraries are strictly identical. In Eq.\ref{cosine}, we use the subscript $CF$ for collaborative filtering, as this quantity is good candidate for measuring the similitude of users depending on their profiles. 

In the following, we also look at the correlations $C_{CF}^{ij}$, that measures the correlations of groups depending on their common audiences, and $C_{CT}^{IJ}$ ($CT$ for collaborative tagging) that measures the correlations of the tags depending on the items to which they are attributed.

\subsubsection{Structure analysis}

\begin{figure}
\hspace{1.5cm}
\includegraphics[width=6.1in]{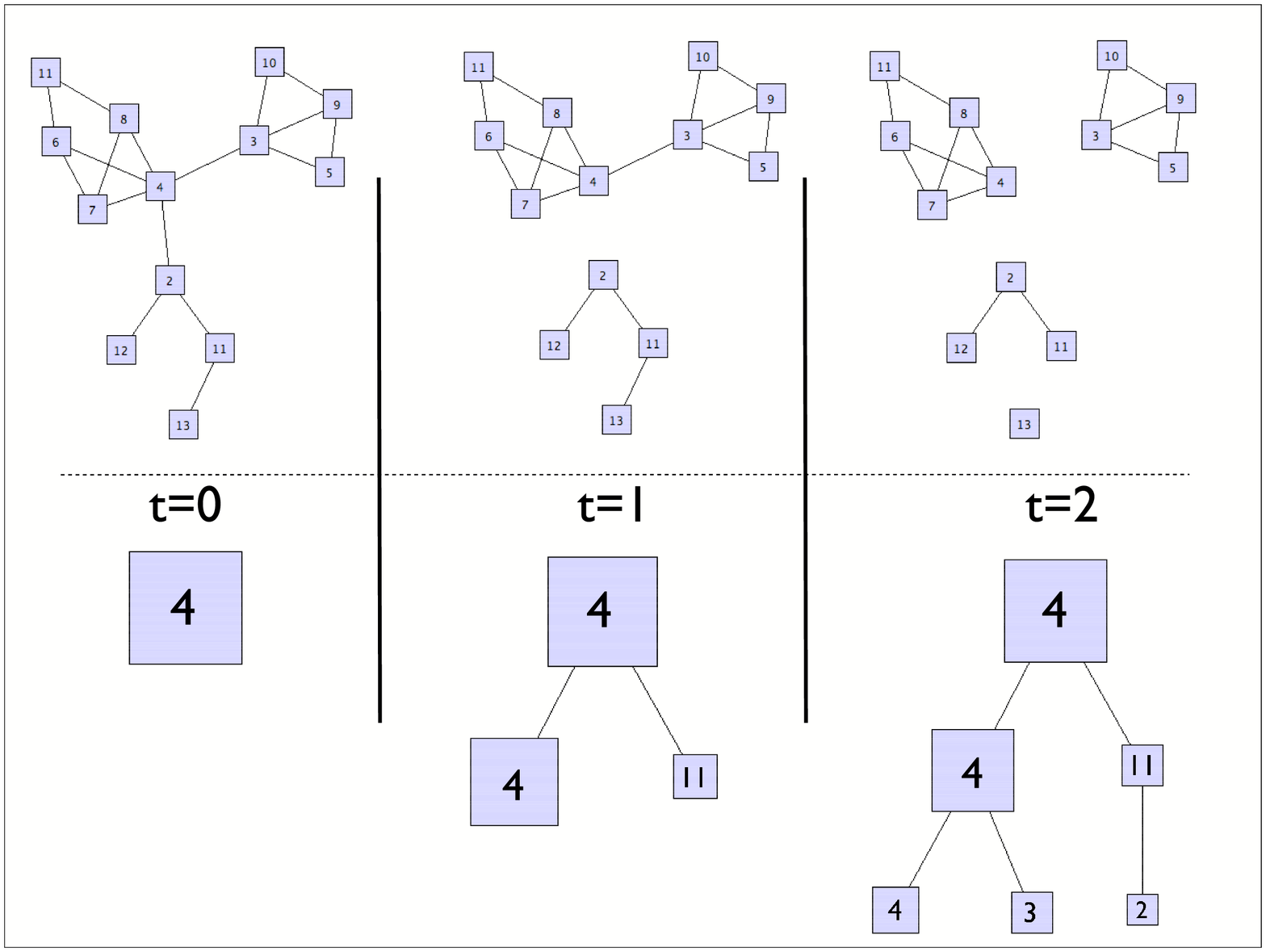}
\caption{\label{explication}  Branching representation of a squared correlation matrix of 13 elements. At each increasing step (t=0,1,2) of the filter $\phi$, links are removed, so that the network decomposes into isolated islands. These islands are represented by  squares, whose size depends on the number of nodes in the island.  Starting from the largest island, branches indicate a parent relation between the islands. Moreover, we indicate one characteristic item for each node of the tree representation. To do so, we look at all the elements belonging to the corresponding island, and chose the item $i$ that maximises $\sum_j C^{ij}$, where the sum is performed over all items of the island. In this explicative case, it is the item 4 at $t=0$, the items $4$ and $11$ at $t=1$...
}
\end{figure}

 At this level, the search for structures requires the analysis of large correlation matrices, and the  uncovering of connected blocks that could be identified as families/genres/communities. 
 In order to extract families of alike elements from the  correlation matrix $\bf C$, we define the filter coefficient $\phi \in [0,1[$ and filter the matrix elements so that $C_{\phi}^{ij}=1$ if $C^{ij}> \phi$ and $C_{\phi}^{ij}=0$ otherwise. 
Starting from $\phi=0.0$, namely a fully connected network, increasing values of the filtering coefficient remove less correlated links and lead to the shaping of well-defined islands, completely disconnected from the main island. 

\begin{figure}

\includegraphics[width=3.1in]{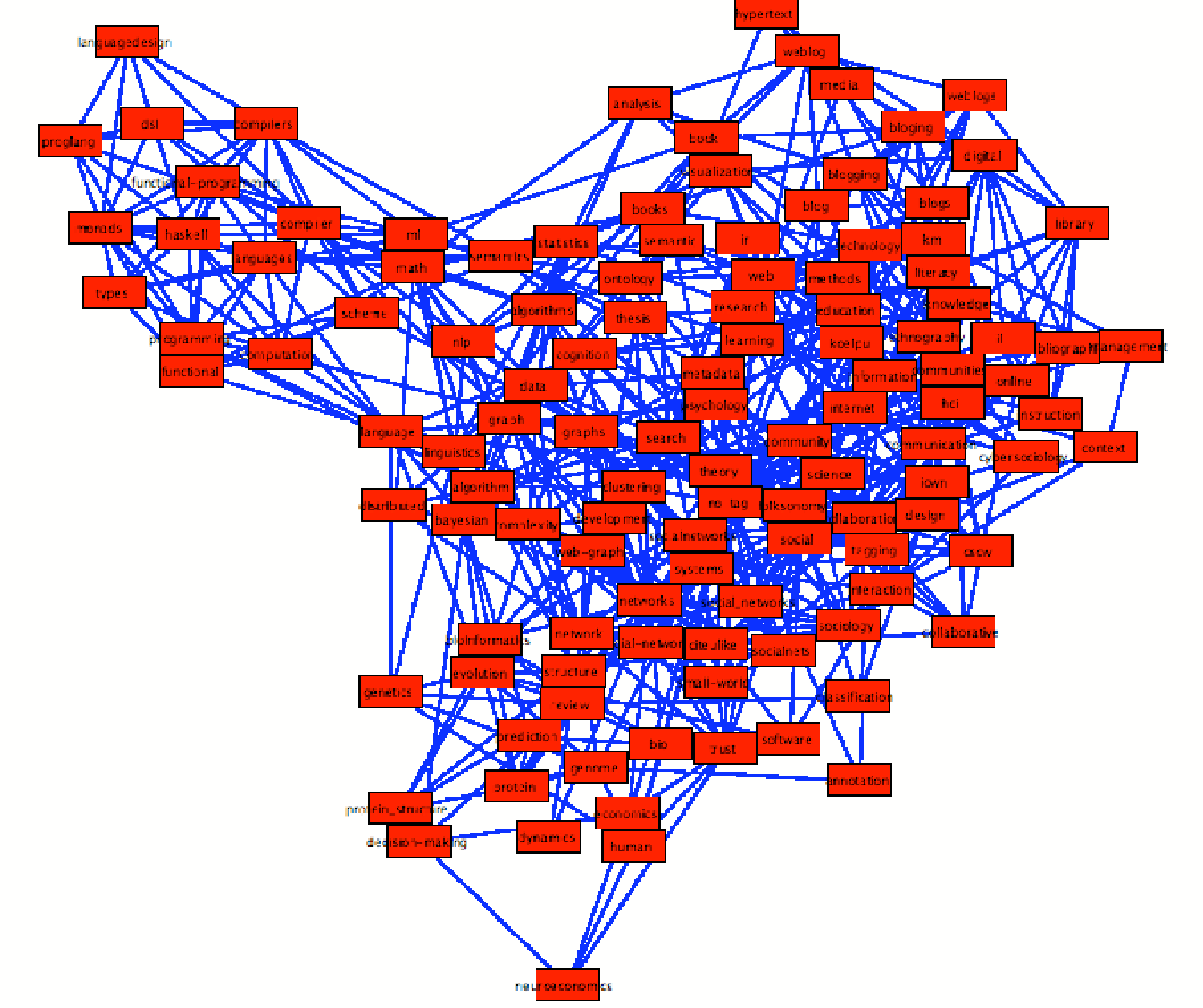}
\includegraphics[width=3.1in]{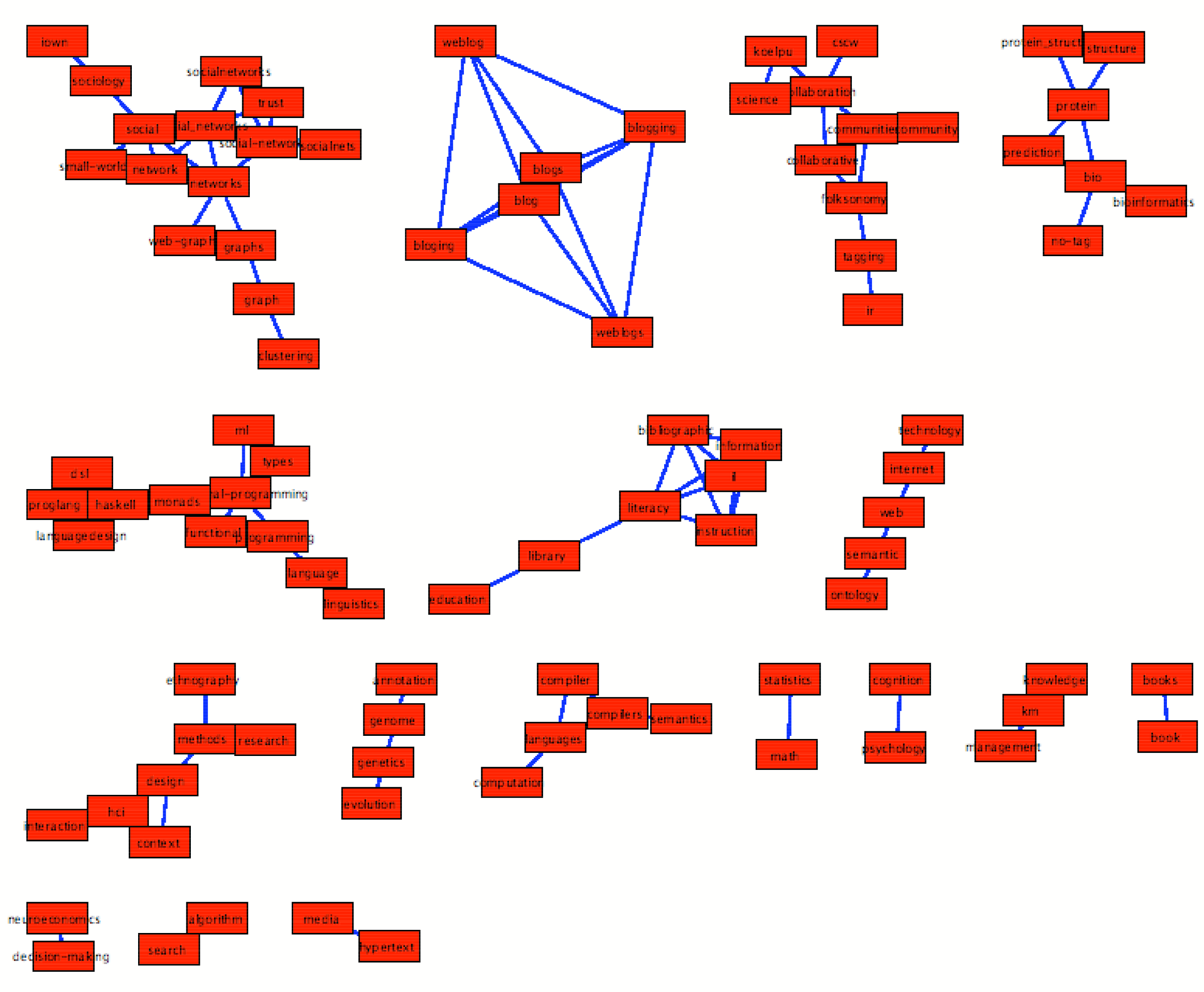}

\caption{\label{perco}  Graph representation of the tags correlation matrix obtained from {\em www.citeulike.com}. Only the 120 most used tags have been considered in the dataset. The values of the filter parameter $\phi$ are 0.1 (before the percolation transition) and 
$\phi$ = 0.25 (for which the system has decomposed into disconnected islands). The 
graphs were plotted thanks to the visone graphical tools \cite{visone}.
}
\end{figure}

\begin{figure}
\hspace{0.5cm}
\includegraphics[width=5.1in]{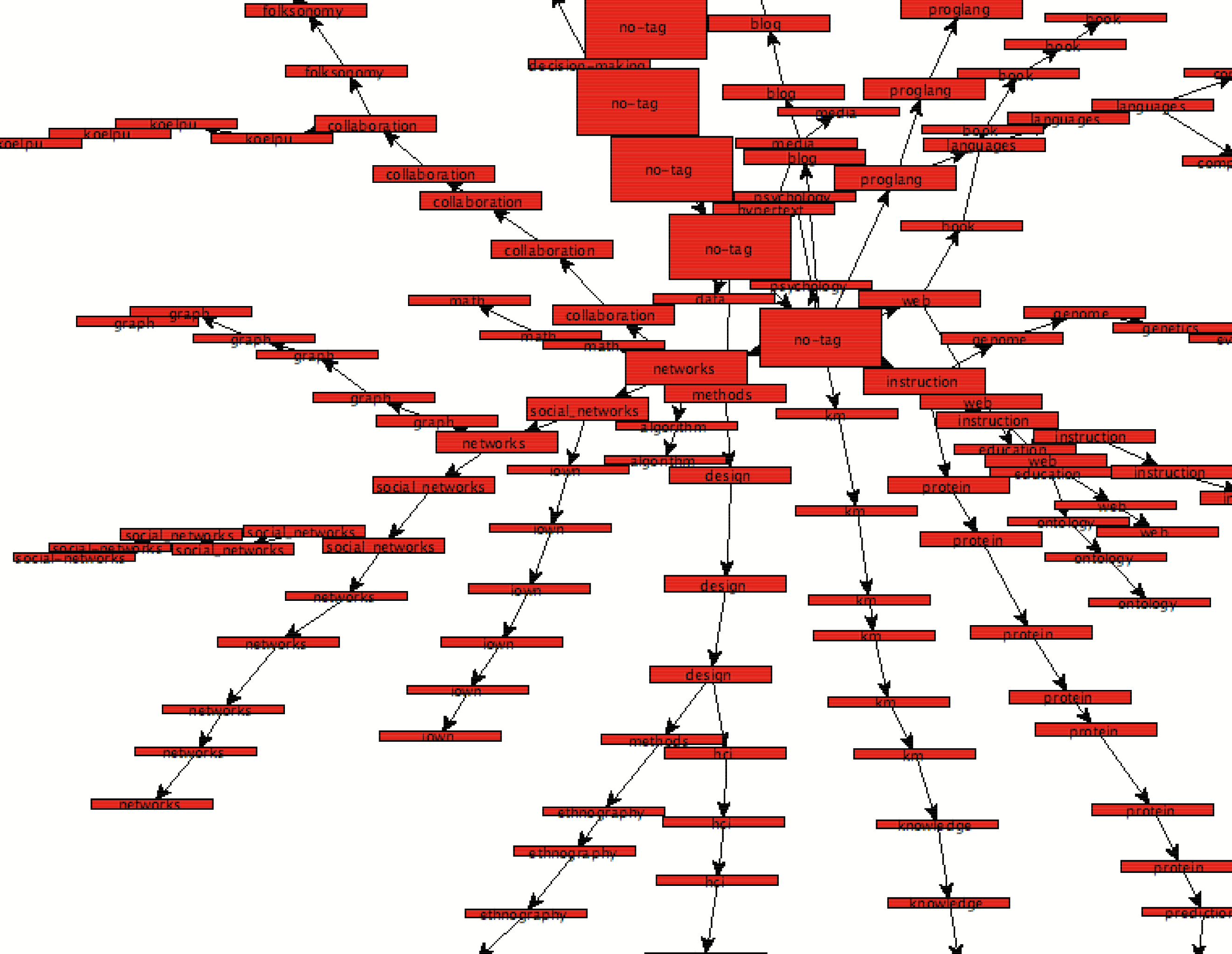}
\caption{\label{trreCite}  Branching representation of the correlation matrix represented in Fig.\ref{perco}.
The filtering, with parameter ranging from 0.1 to 0.4 induces 
a snake of squares at each filtering level. The shape of the 
snake as well as its direction are irrelevant.
}
\end{figure}

A branching representation of the community structuring \cite{lambi} is used to visualise the process (see Fig.\ref{explication} for the sketch of three first steps of an arbitrary example). To do so, we start the procedure with the lowest value of $\phi=0.0$, and we represent each isolated island by a square whose surface is proportional to its number of internal elements. Then, we increase slightly the value of $\phi$, e.g. by 0.05, and we repeat the procedure.  From one step to the next step, we draw a bond between emerging sub-islands and their parent island. The filter is increased until all bonds between nodes are eroded (that is, there is only one node left in each island). Let us note that islands composed of only one element are not depicted for the sake of clarity. Applied to the above correlation matrix $C^{ij}$, the  tree structure gives some insight into the specialisation by following branches from their source (top of the figure) toward their extremity (bottom of the figure).

By construction, the above procedure unambiguously attributes  to each element a hierarchical set of categories. 
Consequently, starting from collaborative filtering that is a non-exclusive and non-hierarchical process, we have arrived to an exclusive and hierarchical structure that may be viewed as a taxonomy. This relation could have helpful applications in order to automatically structure content in systems without a central authority. 
 
\subsection{Data Analysis}

This work is based on the analysis of data retrieved from collaborative filtering websites. We detail in the following the data obtained from each site.

\subsubsection{www.audioscrobbler.com}

A database has been downloaded from {\em audioscrobbler.com} in January 2005. It consists of a listing of  users (each represented by a number), together with the list of music groups that the users own in their library.  
In the original data set, there are  $617900$ different music groups and $35916$ users.  On average, each user owns 140 music groups in his/her library, while each group is owned by 8 persons. 
An analysis has been performed on a subset of the top 1000 most-owned groups \cite{lambi}. 
The resulting tree representation exhibits long persisting branches, some of them leading to standard, homogenous style groupings, while many other islands are harder to explain from a standard genre-fication point of view. In order to complete the data, we have also downloaded  from $http://www.lastfm.com$ a list of the  genres tagged by people on music groups, together with the number of times this description occurred, thereby empirically measuring $\gamma^i_{|_{\alpha}}$ and $C_{CT}^{IJ}$. We refer the reader to the original papers \cite{lambi,lambi2} for a detailed analysis of the dataset.

\subsubsection{www.citeulike.org}

Automatic download in August 2005 has allowed to retrieve a database from $citeulike$.  It consists of a list of articles, readers and tags. In the database, there are 120435 articles, $4453$ users and $36377$ different tags, the most important being in decreasing order: {\em review, evolution, theory, multiagent,  networks, statistics, learning, thesis, history...}. It is interesting to note that there are no mainstream articles in the system. Indeed, the most owned article is {\em Semantic blogging and decentralized knowledge management}, by S. Cayzer, but its audience only includes 40 people, i.e. less than one percent of the total audience.
This behaviour is in opposition with what occurs in music \cite{lambi}, where some music groups like $Radiohead$ may be owned by a majority of people, and where the audience of groups has been shown to behave like a power-law. This very high degree of specialisation observed in science should require a modelling that accounts for the specific tastes and interests of scientists, and that goes beyond the usual preferential attachment mechanism \cite{albert}. In Fig.3 and Fig.4, we plot the observed decomposition of the matrix  $C_{CT}^{IJ}$ of the top 120 tags.
Finally, let us stress that the use of collaborative filtering methods to science 
opens perspectives in scientometrics, by proposing a science classification based on the audience of papers, instead of the use of keywords or citations. 

\section{Applications: measuring diversity}

\begin{figure}
\hspace{1.5cm}
\includegraphics[width=5.2in]{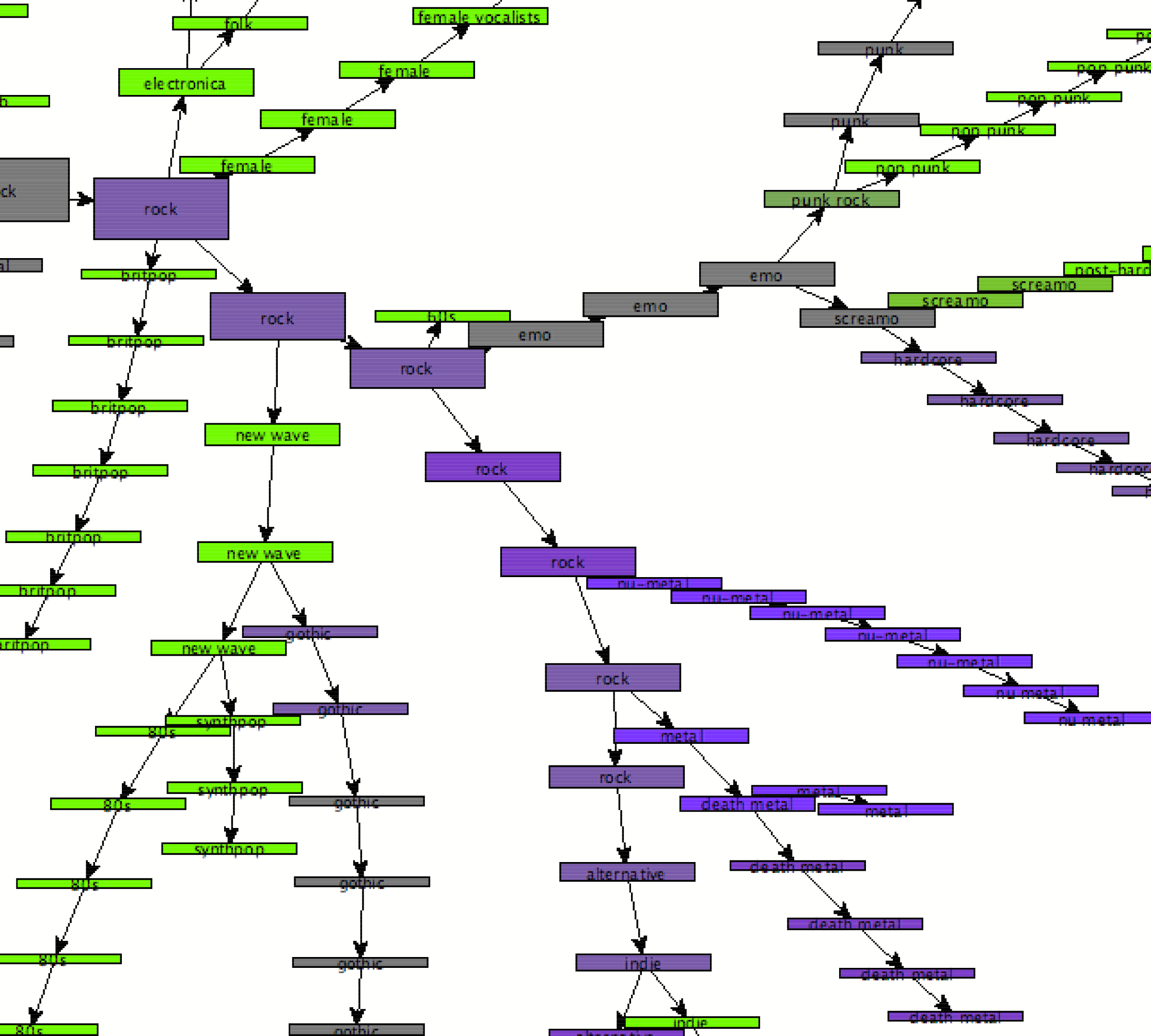}
\caption{\label{explication2}  Diversity representation (see section ${\bf IV.A}$) of one particular user, based on the tree representation of the whole set of users. The analysis is performed on the data obtained from {\em audioscrobbler} and {\em last.fm}. The considered person is  very {\em Rock-oriented} and listens mainly to {\em metal} and {\em death metal}.}
\end{figure}

In the above sections,  methods have been introduced in order to uncover the structures that arise from a collaborative description of content. 
A first application has been outlined above, namely the possibility  to construct from a decentralised classification of content an objective hierarchical taxonomy. In this last section, we discuss another interesting application of this work, i.e. the possibility to compare the tastes and interests of different persons, as well as the measure of their {\em diversity}. For the sake of clarity, we focus in the following on the case of music collaborative websites. But the results are directly applicable to any collaborative systems.

Practically, let us consider the case of two users $\mu_1$ and $\mu_2$ who own a list of music groups, each of them characterised by a spectrum of genres. From this knowledge, one would like to have a quantitative measure of the diversity of the persons, as well as a way to measure whether they have a similar taste. Let us note ${\mbox{\boldmath$\tau$}}_{\mu_1}$ and ${\mbox{\boldmath$\tau$}}_{\mu_2}$ the vector of genres characterising $\mu_1$ and $\mu_2$, where  
\begin{equation}
{\mbox{\boldmath$\tau$}}_{\mu_1}= (\tau_{\mu_11}, ..., \tau_{\mu_1I}, ..., \tau_{\mu_1n_T}).
\end{equation}
$ \tau_{\mu_1I}$
 is the number of times that the tag $I$ is associated to an item of $\mu_1$ and $n_T$ is the total number of tags in the system. A naive way to study diversity consists in implicitly assuming that all tags have different meaning and in characterising a person by the width of the distribution of $\tau$. This is what we have done in ref.\cite{lambi2}, where we defined a probabilistic entropy in order to measure these fluctuations.
It is nonetheless an oversimplification that does not take into account the correlations between the tags, i.e. the fact that tags may have more or less equivalent meanings.

\subsection{Colour-based visualisation of diversity}
A more refine measure of diversity should 
require a proper counting of the {\em categories} to which the user belongs.
To do so, we propose to visualise the
branches and sub-branches of the hierarchical tree (Fig.2) in which the user is more active than the average.  Let us assume that, at some level of the filtering, an island (the node of one branch in the tree representation) is composed of $K$ tags, say $I_1, ..., I_i, ..., I_K$. Let us denote $\tau_{I_i}^S$ the total number of times the tag $I_i$ is used in the sample, while, as defined above, $\tau_{\mu I_i}$ is the total number of times  $I_i$ is tagged to the items belonging to $\mu$.
The above island, composed of $K$ genres, is then characterised by:

- $p^S = (\sum_{i=1}^K \tau_{I_i}^S)/ (\sum_{I=1}^{n_T} \tau_{I}^S)$, that gives the empirical probability that a tag used in the sample belongs to the considered island.

- $p^{\mu} = (\sum_{i=1}^K \tau_{I_i}^\mu)/ (\sum_{I=1}^{n_T} \tau_{I}^\mu)$, that is the probability that a tag used on an item of $\mu$ belongs to the same island.

The activity of the user in the island is simply evaluated by looking at the ratio $r=p^{\mu}/p^{S}$. By construction, this quantity is bigger than 1 if the user owns many groups belonging to this island, and smaller than 1 otherwise. In Fig.5, we apply the method to all the nodes of the tree representation, and use a colour representation in order to represent the value of $r$, i.e. the nodes are printed in a colour ranging from green (low $r$) to blue (high $r$).  
One should also note that the above method allows a rapid comparison between different users, by looking whether they are active in the same branches or in different branches.

\subsection{Metric approach}
From a quantitative point of view, a first approximation consists in averaging {\em distances} between all possible pairs of  genres. To do so, we define the sine matrix $\bf S$ associated to the $n_T \times n_T$ matrix $\bf C$, where $S_{IJ}=\sqrt{1-C^2_{IJ}}$. Using this matrix as a metric, we calculate the diversity of a person $\mu$ with 
\begin{equation}
d_{\mu}=\sum_{IJ} S_{IJ} \tau_{\mu I} \tau_{\mu J}.
\end{equation}
 The higher this value, the more the person is described by a large number of relevant {\em orthogonal} terms.
Accordingly, the equivalence between two persons is measured by measuring the average distance between their tags, together with a proper normalisation:  
$d_{\mu_1 \mu_2}=\sum_{ij} S_{IJ} \tau_{\mu_1 I} \tau_{\mu_2 J}/\sqrt{d_{\mu_1}  d_{\mu_2}}$. Preliminary results based on the empirical data of {\em citeulike} and {\em audioscrobbler} seem to confirm the better description obtained by using these distances, instead of the entropy description of \cite{lambi2}.

\begin{acknowledgments}
Figures  1, 2, 3, 3 and 5 were plotted thanks to the {\em visone} graphical tools \cite{visone}.
R.L. would like to thank especially  G. D'Arcangelo and A. Scharnhorst for fruitful discussions.
This work 
has been supported by European Commission Project 
CREEN FP6-2003-NEST-Path-012864.
\end{acknowledgments}


\begin{thebibliography}{0}
%
\bibitem{lambi}
R. Lambiotte and M. Ausloos,  {\em Phys. Rev. E}, {\bf 72}, 066107 (2005)
%
\bibitem{lambi2}
R. Lambiotte and M. Ausloos, {\em EPJB}, in press; physics/0509134
%
\bibitem{gold}
S. Golder and B. A. Huberman, cs.DL/0508082
%
\bibitem{deus} 
For instance, see the interview of the Belgian band {\em dEUS}, on $http://www.metroactive.com/$ $papers/metro/07.17.97/deus-9729.html$,
whose diverse influences include  {\em ABBA, Sonic Youth, Captain Beefheart, Franck Zappa...}
\bibitem{visone}
{\em http://www.visone.de/}
\bibitem{albert}
R. Albert and A.L. Barab\'asi, {\em Rev. of Mod. Phys.}, {\bf 74} (2002) 47
%
\end{thebibliography}
\end{document}